\documentclass[aps,prl,nofootinbib,10pt,tightenlines,notitlepage, twocolumn,floatfix,eqsecnum,superscriptaddress,showkeys]{revtex4-1}
    
\usepackage[utf8]{inputenc}          
\usepackage{graphicx}         
\usepackage[table]{xcolor}
\usepackage{array,dcolumn,longtable} 
\usepackage{amsmath,amssymb,amsfonts,slashed} 
\usepackage{mathtools}          

\usepackage[linktocpage,breaklinks]{hyperref}

\usepackage{txfonts}
\usepackage{bm}
\usepackage{stmaryrd}
\usepackage{tensor}
\usepackage[utf8]{inputenc}

\usepackage{epsfig}
\usepackage{epstopdf}

\usepackage{natbib}
\usepackage{cleveref}

\usepackage{color}
\usepackage{colortbl}
\definecolor{midgreen}{rgb}{0.52, 0.73, 0.4}

\definecolor{purple}{rgb}{0.52, 0., 0.52}

\def\OO{$^{16}\mathrm{O}^{16}\mathrm{O}$ }

\begin{document}

\title{\OO collisions at RHIC and the LHC comparing $\alpha$ clustering vs substructure}
\date{\today}

\author{Nicholas Summerfield}
\affiliation{Department of Physics, University of Houston, Houston, TX 77204, USA}
\author{Bing-Nan Lu}
\affiliation{China Academy of Engineering Physics, Graduate School,
Building 8, No.~10 Xi'er Road, ZPark II, Haidian District, Beijing, 100193,
China}
\author{Christopher Plumberg}
\affiliation{Illinois Center for Advanced Studies of the Universe, Department of Physics, University of Illinois at Urbana-Champaign, Urbana, IL 61801, USA}
\author{Dean Lee}
\affiliation{Facility for Rare Isotope Beams and Department of Physics and Astronomy,
Michigan State University, East Lansing, MI 48824, USA}
\author{Jacquelyn Noronha-Hostler}
\affiliation{Illinois Center for Advanced Studies of the Universe, Department of Physics, University of Illinois at Urbana-Champaign, Urbana, IL 61801, USA}
\author{Anthony Timmins}
\affiliation{Department of Physics, University of Houston, Houston, TX 77204, USA}

\begin{abstract}
Collisions of light and heavy nuclei in relativistic heavy-ion collisions have been shown to be sensitive to nuclear structure.  With a proposed \OO run at the LHC and RHIC we study the potential for finding $\alpha$ clustering in $^{16}$O.  Here we use the state-of-the-art iEBE-VISHNU package with $^{16}$O nucleonic configurations from {\rm ab initio} nuclear lattice simulations.  This setup was tuned using a Bayesian analysis on pPb and PbPb systems.  We find that the \OO system always begins far from equilibrium and that at LHC and RHIC it approaches the regime of hydrodynamic applicability only at very late times. Finally, by taking ratios of flow harmonics we are able to find measurable differences between $\alpha$-clustering, nucleonic, and subnucleonic degrees of freedom in the initial state.

\end{abstract}
\maketitle



\noindent {\bf Introduction} In the past several years, the state-of-the-art in the field of relativistic nuclear collisions has reached the threshold of precision physics \cite{Noronha-Hostler:2015dbi,Niemi:2015voa, Adam:2015ptt,Bernhard:2019bmu,Shen:2020mgh,Bernhard:2019bmu}.  The evolution of nuclear collisions is by now widely accepted to be  well-described within the framework of relativistic hydrodynamics, in which fluid dynamical behavior is manifested by a collective response to the initial collision geometry \cite{Teaney:2010vd,Gardim:2011xv,Niemi:2012aj,Teaney:2012ke,Qiu:2011iv,Gardim:2014tya,Betz:2016ayq,Rao:2019vgy,Hippert:2020kde}.  Precision measurements for probing the hydrodynamic evolution of nuclear collisions include a suite of flow observables \cite{Schukraft:2012ah,Acharya:2017gsw}, multi-particle correlation observables \cite{Bilandzic:2010jr,Luzum:2013yya,Zhou:2016eiz}, soft-hard/heavy multiparticle azimuthal correlations \cite{Betz:2016ayq,Prado:2016szr,Acharya:2018bxo,Katz:2019qwv,CMS:2021rlx}, and femtoscopic radii \cite{Kisiel:2008ws, Adamczyk:2014mxp, Adare:2015bcj, Adam:2015vna, Plumberg:2020jod}, to name a few. 

These observables, moreover, are sensitive to a variety of stages in nuclear collision evolution, including the initial state, the pre-hydrodynamic evolution \cite{Liu:2015nwa,NunesdaSilva:2020bfs,Schenke:2019pmk,Giacalone:2020byk}, and the subsequent hydrodynamic  phase.  Essential to disentangling the effects of quantum fluctuations in the initial state and the pre-hydrodynamic evolution from those of the subsequent medium response is the ability to engineer initial conditions with specified geometries.  This approach has been exploited already with great success in the context of small-system geometry engineering by the PHENIX collaboration \cite{	Aidala:2018mcw,	Adare:2018toe,	Adare:2015ctn,	Aidala:2016vgl,	Adare:2017wlc,	Adare:2017rdq,	Aidala:2017pup,	Aidala:2017ajz}.  More recently, dedicated runs of \OO collisions have been proposed \cite{Citron:2018lsq, Sievert:2019zjr, Katz:2019qwv, Huang:2019tgz, Rybczynski:2019adt, Huss:2020whe, Schenke:2020mbo, Brewer:2021kiv} at both RHIC and LHC as a way of extending the geometry scan results to systems of intermediate size, which exhibit more exotic initial configurations due to an effect known as ``$\alpha$-clustering".

The phenomenon of $\alpha$-clustering is a type of nucleon-nucleon (NN) correlation which is expected on the basis of nuclear lattice effective field theory (NLEFT) calculations to be present in doubly magic nuclei such as $^{16}\mathrm{O}$ and $^{208}\mathrm{Pb}$.  In such nuclei, nucleon positions are not completely uncorrelated, but tend to cluster together into groupings of two neutrons and two protons each, thereby effectively forming $\alpha$ particles (or ``$\alpha$ clusters") in the nucleus.  These correlations lead to quantifiable effects on the initial states of collisions between such nuclei and may manifest themselves in corresponding precision measurements of nuclear collision flow observables \cite{Rybczynski:2019adt, PhysRevC.102.054907}.  It may also be possible to have subnucleonic fluctuations that would influence the collective flow \cite{Moreland:2012qw,Dumitru:2014yza,Noronha-Hostler:2015coa,Albacete:2017ajt,Mantysaari:2017cni,Gardim:2017ruc,Moreland:2018gsh}.  A natural question is thus whether $\alpha$-clustering is measurable in relativistic heavy-ion collisions once all relevant effects have been considered. 

The purpose of this paper is to explore the quantitative impact on flow observables of incorporating $\alpha$-clustering effects vs. subnucleonic fluctuations into the initial conditions for hydrodynamic simulations of \OO at both RHIC and LHC energies. To do this we adopt the state-of-the-art setup used in a recent Bayesian analysis \cite{Bernhard:2019bmu} which was conditioned on experimental data at the LHC.

\noindent {\bf Initial Conditions} 
A typical heavy ion collision is rarely head on, but is rather characterized by a finite impact parameter.  Consequently, a certain number of nucleons do not participate in the collision and simply travel on to the detector.  The nucleons that do collide are counted using $N_{part}$ and the impact region is treated as the initial condition for relativistic hydrodynamic calculations. In recent years \cite{Broniowski:2013dia,Adamczyk:2015obl,wang:2014qxa,Moreland:2014oya,Goldschmidt:2015kpa,Giacalone:2017dud,Rybczynski:2017nrx,Schenke:2019ruo,CMS:2018jmx,Acharya:2018ihu,ATLAS:2018iom} it has been found that the shape of the nucleus can play a role in the geometrical shape of the impact range, which is quantified through eccentricities. These eccentricities are connected to the collective flow observables through linear response for central \cite{Teaney:2010vd,Gardim:2011xv,Niemi:2012aj,Teaney:2012ke,Qiu:2011iv,Gardim:2014tya,Betz:2016ayq} and mid-central collisions and linear+cubic response in peripheral collisions \cite{Noronha-Hostler:2015dbi}. Thus, deformations in the shape of the nucleus are then translated to final state observables, which are most detectable in central collisions from linear response. 



The simplest approximation for initial collisions in relativistic heavy-ion collisions is to treat each nucleus as a two-parameter Woods-Saxon density distribution in the nuclear rest frame, written in spherical coordinates as:
\begin{align} \label{e:2PF}
\rho(r,\theta,\phi) = \rho_0 \left[ 1 + \exp\left(\frac{r - R}{a}\right)\right]^{-1}
\end{align}
with $\rho_0$ the nuclear saturation density, $R$ a measure of the gluonic radius of the nucleus, and $a$ the surface diffusion parameter.  For some nuclei a three-parameter generalization \cite{DeJager:1987qc} of the nuclear density has been extracted instead of (or in addition to) the standard Woods-Saxon distribution \eqref{e:2PF}.  This three-parameter fit modifies the radial density distribution somewhat:
\begin{align} \label{e:3PF}
\rho(r,\theta,\phi) = \rho_0 \left( 1 + w \frac{r^2}{R^2} \right)
\left[ 1 + \exp\left(\frac{r - R}{a}\right)\right]^{-1} .
\end{align}
For nuclei such as ${}^{208}$Pb, a ``doubly magic'' nucleus in the nuclear shell model, these spherically symmetric densities give a good description of elliptic flow at the LHC.  
\begin{table}
\begin{tabular}{|c|c|ccc|} 
	\hline
			& Parameterization	& $R$ (fm)	& $a$ (fm)	& $w$ (fm)		\\
	\hline
	${}^{16}$O & 3pF	& 2.608		& 0.513		& -0.051	 \\
	\hline
\end{tabular}
\caption{Parameters for the Wood-Saxon density distribution used in the initial conditions.}
\label{t:params}
\end{table}
The parameters used in our initial conditions are given in Table~\ref{t:params}.  For $^{16}$O, only the three-parameter fit \eqref{e:3PF} is available. Being a doubly magic nucleus, $^{16}$O is taken to be spherically symmetric.  We have coded this Wood-Saxon into the phenomenologically driven initial condition model, T$_{\rm R}$ENTo \cite{Moreland:2014oya}, using the following parameters: the thickness function scaling $p=0$, the multiplicity fluctuations $k=1.6$, the nucleon width $\omega=$ 0.51 fm. The nucleon-nucelon cross-sections correspond to the p-p values at each energy investigated: $\sigma_{\rm NN} = 42.5$ mb at $\sqrt{s_{\rm NN }} = 200 $ GeV (RHIC), and $\sigma_{\rm NN} = 72.5$ mb at $\sqrt{s_{\rm NN }} = 6.5 $ TeV (LHC). 

\noindent {\bf \textit{Ab initio} structure and clustering}  Nuclear clustering is a feature of many light nuclear systems and is particularly prevalent in nuclei with even and equal numbers of protons and neutrons.  For such nuclei the clustering is mostly associated with the formation of $\alpha$ clusters.  See, for example, Ref.~\cite{Freer:2017gip} for a recent review.  The nuclear states with the most pronounced $\alpha$ cluster substructures are excited states near $\alpha$ separation thresholds, such as the Hoyle state of $^{12}$C.  However the strong four-nucleon correlations also persist in ground states of nuclei. Recently it has even been suggested that the parameters of the nuclear force lies close to quantum phase transition between a nuclear liquid and a Bose gas of $\alpha$ particles \cite{Elhatisari:2016owd}.  

One of the {\it ab initio} methods that is able to probe $\alpha$-clustering is NLEFT.  See Ref.~\cite{Lee:2008fa} and \cite{Lahde:2019npb} for reviews. 
In this work we use the nucleonic configurations for $^{16}$O produced in Ref.~\cite{Lu:2018bat}.  These calculations used a simple leading order interaction, although the reproduction of the binding energies and radii of light and medium mass are accurate to a few percent error.  In particular, the charge density distribution for $^{16}$O is in excellent agreement with electron scattering data.  These calculations were performed with a $1.32$ fm spatial lattice spacing.

The nucleon configurations were computed using the pinhole algorithm introduced in Ref.~\cite{Elhatisari:2017eno}.  The pinhole algorithm produces a classical distribution of the nucleon positions weighted according to the $16-$nucleon density correlation function for the $^{16}$O ground state.  These $16-$nucleon configurations provide the initial conditions for our hydrodyanmics calculations to be described below.

%






\noindent {\bf Hydrodynamic Setup} 
We model the hydrodynamic evolution in \OO using the Duke Bayesian tune of the iEBE-VISHNU package \cite{Moreland:2018gsh, Bernhard:2019bmu} to p--Pb and Pb--Pb collisions at the LHC.  The framework uses the T$_{\rm R}$ENTo model \cite{Moreland:2014oya} to generate an initial entropy distribution. These distributions require normalization constants of 5.3 (RHIC) and 17 (LHC), which were obtained from an extrapolation of the energy dependence elsewhere \cite{Bernhard:2019bmu}. The initial entropy distribution is then passed through a free-streaming phase of duration $\tau_s = 0.37$ fm$/c$ and then used to initialize the hydrodynamic evolution at $\tau = \tau_s$.  The construction of the hydrodynamic equation of state, as well as the temperature dependences of the specific bulk and shear viscosities $(\zeta/s)(T)$ and $(\eta/s)(T)$ are described in Ref.~\cite{Moreland:2018gsh}.  Finally, the hydrodynamic phase is terminated at a freeze-out temperature of $T_{fo} = 151$ MeV, at which point the system is converted to particles and evolved until kinetic freeze-out using UrQMD \cite{Bass:1998ca, Bleicher:1999xi}.  The final output is a collection of discrete particles at a final timestep which may be used to compute observables of interest, such as flow coefficients and their ratios.


In order to make direct comparisons with experimental data, cumulants of the flow harmonics \cite{Bilandzic:2010jr} are calculated using:
\begin{eqnarray}\label{eqn:cumulants}
\nonumber v_n\{2\}^2 &=& \langle v_n^2 \rangle, \\
\nonumber v_n\{4\}^4 &=& 2 \langle v_n^2\rangle^2 - \langle v_n^4\rangle,
\end{eqnarray}
where the moments of the $v_n$ distribution are used to calculate the cumulants. 
Centrality class bins are determined based on the initial state entropy density, which has been found to be very good proxy for final state multiplicity distributions (used for experimental data).We have run 30,000 events for each different ion and configuration, and use sub-sampling to determine statistical error.

\begin{figure}[t]
\centering
\includegraphics[width=\linewidth]{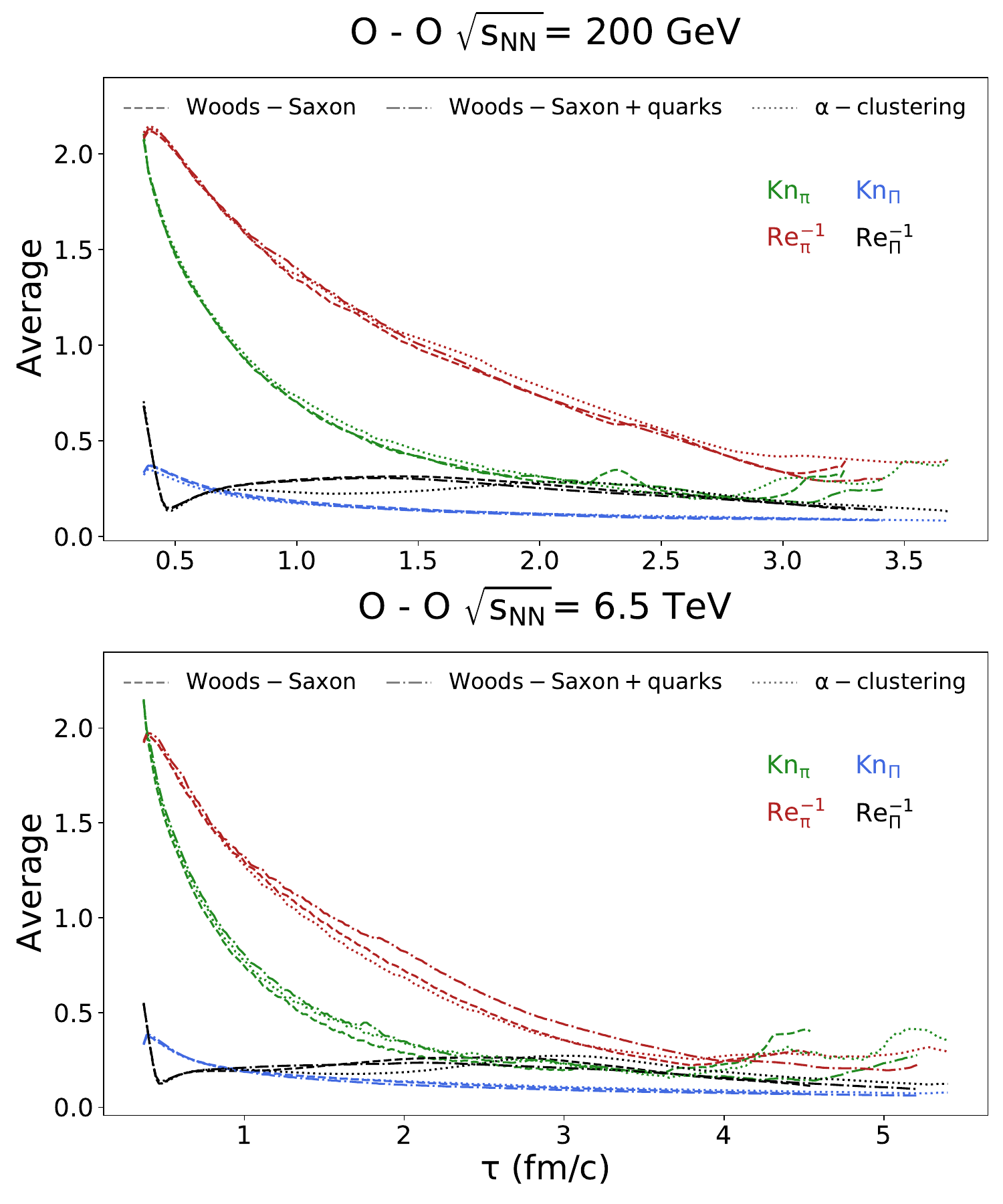}
\caption{(Color online) Average Knudsen and inverse Reynolds values of regions above the freeze out temperature (0.151 GeV) vs time for both $\sqrt{s_{NN}}$ = 200 GeV (top) and $\sqrt{s_{NN}}$ = 6.5 TeV (bottom) comparing the Woods-Saxon, Woods-Saxon + Quarks, and $\alpha$ clustering models with common initial conditions.}
\label{fig:averageKR}
\end{figure}

\noindent {\bf Results}
Hydrodynamics is applicable when there is a large separation of scales.  In relativistic heavy ion collisions there is some ambiguity of the correct scales to compare and, therefore, multiple Knusden $\mathrm{Kn}$ and Inverse Reynolds $\mathrm{Re}^{-1}$ numbers are used \cite{McNelis:2018jho}: 
\begin{align}
	\mathrm{Kn}_\pi &= \tau_\pi \sqrt{\sigma_{\mu\nu}\sigma^{\mu\nu}},
	&\mathrm{Re}_\pi^{-1} &= \sqrt{\pi_{\mu\nu}\pi^{\mu\nu}} / P,\\
	\mathrm{Kn}_\Pi &= \tau_\Pi \theta,
	&\mathrm{Re}_\Pi^{-1} &= \left| \Pi \right| / P
\end{align}

We consider first in Fig.~\ref{fig:averageKR} the time evolution of the $\mathrm{Kn}$ and $\mathrm{Re}^{-1}$ numbers for bulk and shear, averaged at each timestep over all fluid cells above the particlization temperature of $T_{\mathrm{switch}} = 0.151$ GeV for a single event (the same seed for the initial condition is chosen for Wood-Saxon, Wood-Saxon+substructure, and $\alpha$-clustering).   We observe that, while the choice of initial-state model makes little difference to the time-evolution of these quantities, both $\mathrm{Kn}_\pi$ and $\mathrm{Re}^{-1}_\pi$ are problematically large ($\gtrsim 0.5$) for the majority of the hydrodynamic phase, predominantly at early times $\tau \lesssim 2$ fm/$c$.  This observation holds at both RHIC and LHC energies, and suggests that the hydrodynamic formalism is pressed to the limits of its validity in the description of intermediate systems such as \OO. While $\mathrm{Kn}$ and $\mathrm{Re}^{-1}$ have been previously studied in an event averaged version of pPb \cite{Niemi:2014wta}, this is the first study of their values with the setup used within the Duke Bayesian analysis. It appears that eventually reasonable $\mathrm{Kn}$ and $\mathrm{Re}^{-1}$ are reached after $\tau\sim 3$ fm although there is some dependence on both the initial conditions. We note that one must consider the maximum of all $\mathrm{Kn}$ and $\mathrm{Re}^{-1}$ to determine the applicability of hydrodynamics and, therefore, these numbers indicate that even in intermediate systems one needs to consider the implications of far-from-equilibrium effects.

\begin{figure}[h]
\centering
\includegraphics[width=\linewidth]{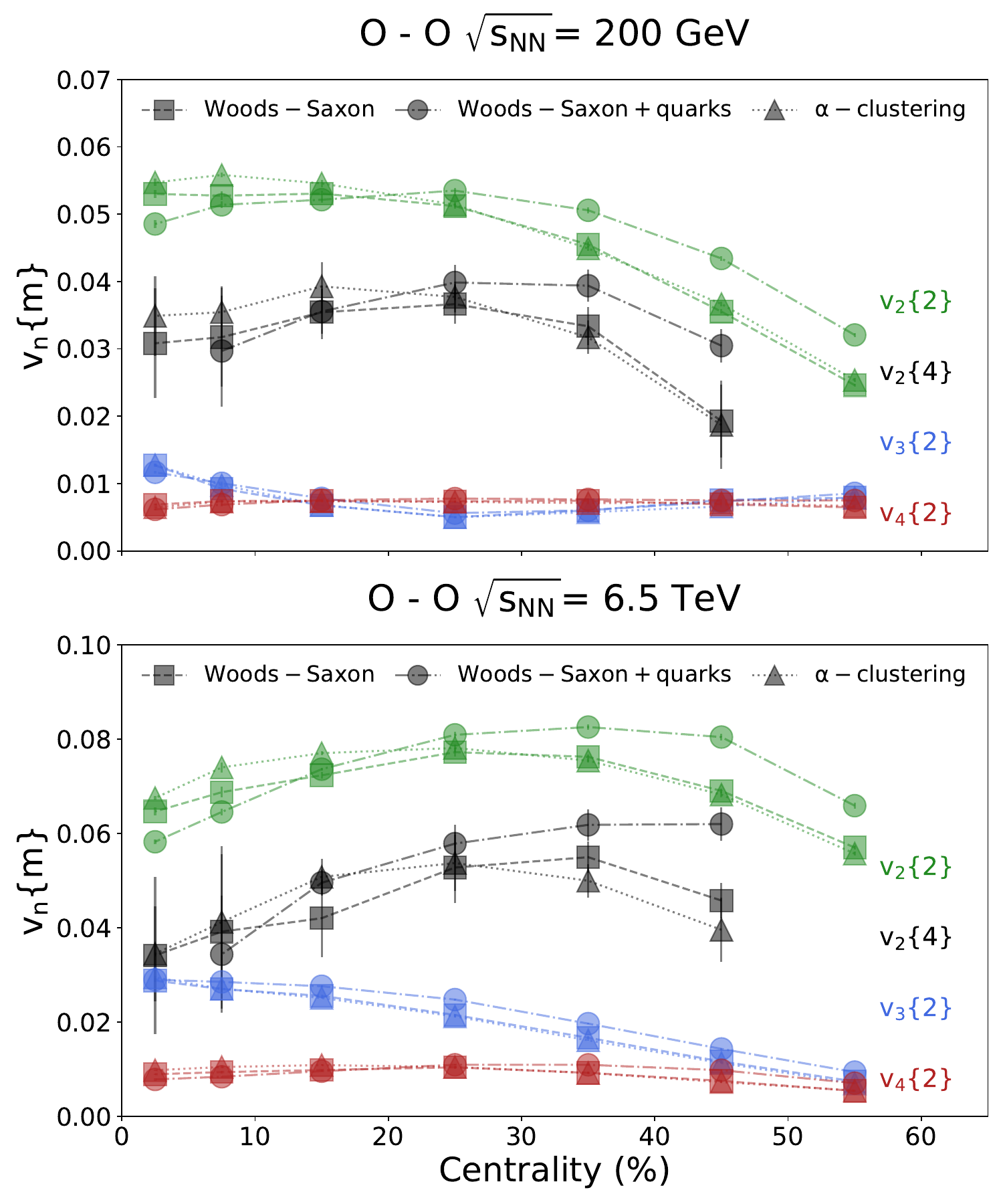}
\caption{(Color online) Various flow coefficients ($v_n\{m\}$) vs centrality for both $\sqrt{s_{NN}}$ = 200 GeV (top) and $\sqrt{s_{NN}}$ = 6.5 TeV (bottom) comparing the Woods-Saxon, Woods-Saxon + Quarks, and $\alpha$ clustering models.}
\label{fig:flowCoefficients}
\end{figure}

In Fig.\ref{fig:flowCoefficients} we present the flow cumulants predicted by our model at RHIC and LHC energies as functions of centrality, for the various initial-state models considered.  We note that the largest quantitative effects are due to subnucleonic fluctuations and emerge at large centralities, while the effects of $\alpha$-clustering are somewhat smaller but on the same order of magnitude, and occur mainly at small centralities.  $v_2$ is the most sensitive to details of the initial state, while $v_3$ and $v_4$ are only weakly affected.   These features are expected from a hydrodynamic response to initial geometry which is dominated by fluctuations in central collisions and by global collision geometry in mid-central and peripheral collisions.

\begin{figure}[h]
\centering
\includegraphics[width=\linewidth]{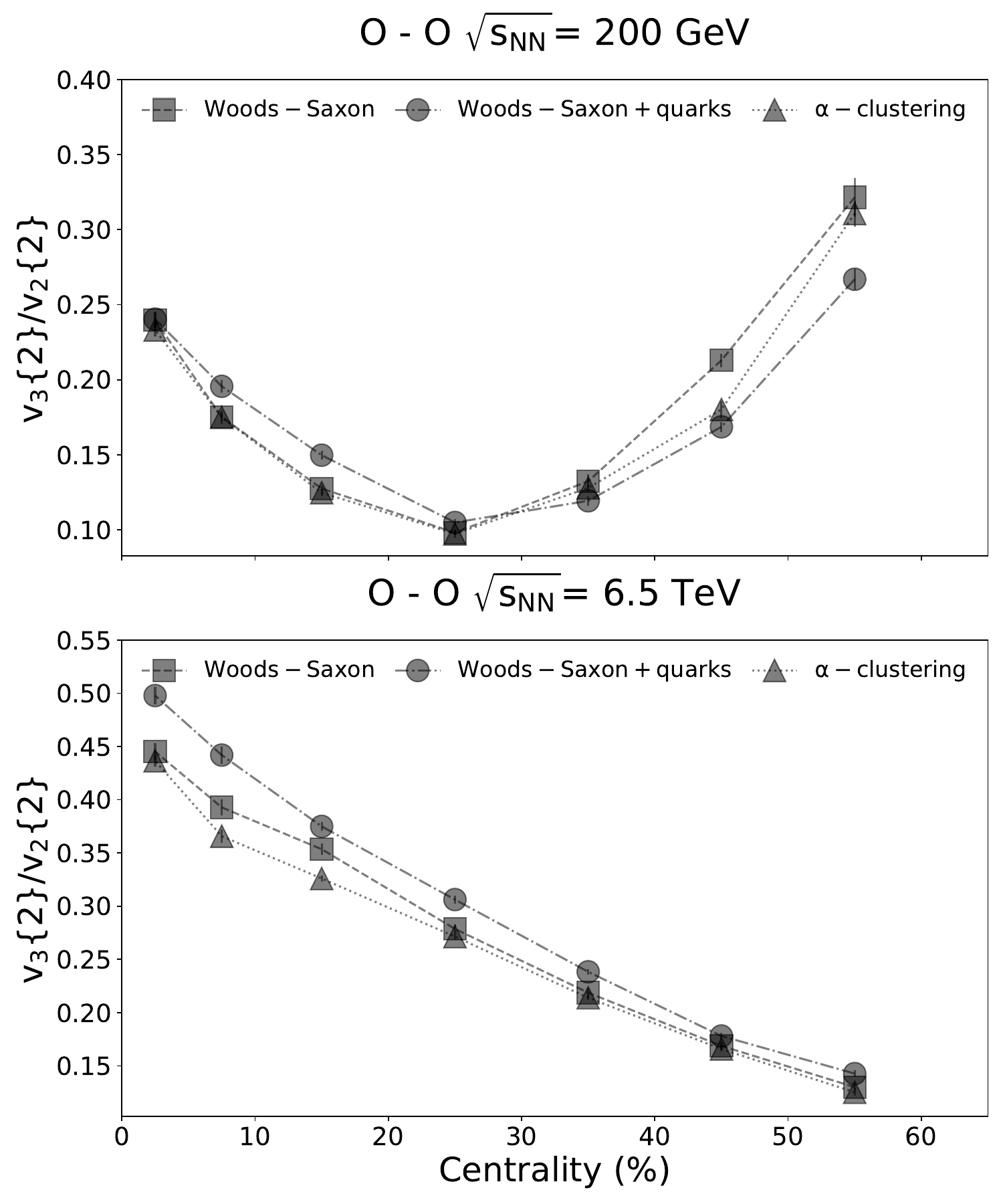}
\caption{$v_3\{2\}/v_2\{2\}$ vs centrality for both $\sqrt{s_{NN}}$ = 200 GeV (top) and $\sqrt{s_{NN}}$ = 6.5 TeV (bottom) comparing the Woods-Saxon, Woods-Saxon + Quarks, and $\alpha$ clustering models.}
\label{fig:fig3}
\end{figure}

\begin{figure}[h]
\centering
\includegraphics[width=\linewidth]{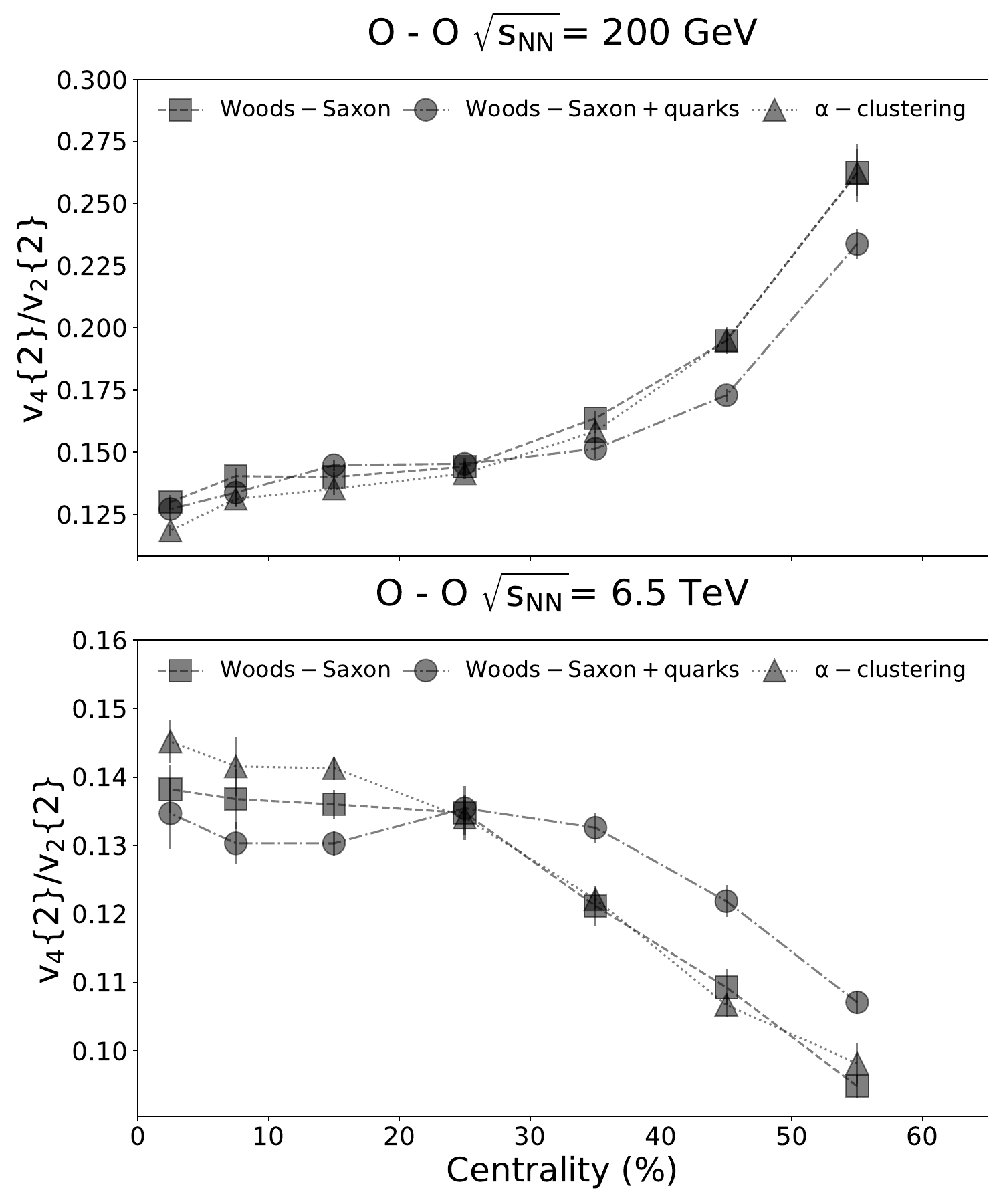}
\caption{$v_4\{2\}/v_2\{2\}$ vs centrality for both $\sqrt{s_{NN}}$ = 200 GeV (top) and $\sqrt{s_{NN}}$ = 6.5 TeV (bottom) comparing the Woods-Saxon, Woods-Saxon + Quarks, and $\alpha$ clustering models.}
\label{fig:fig4}
\end{figure}

\begin{figure}[h]
\centering
\includegraphics[width=\linewidth]{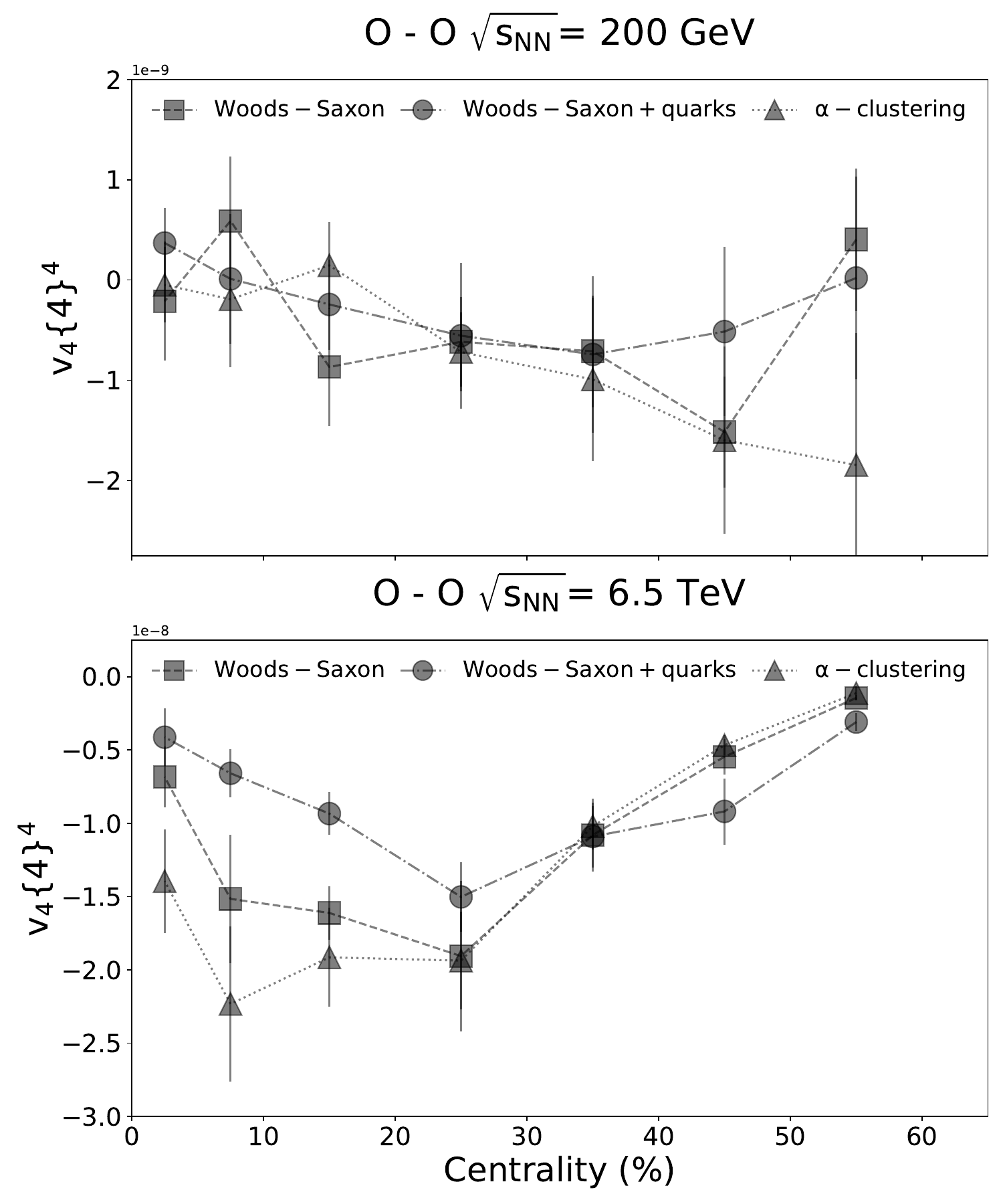}
\caption{$v_4\{4\}^4$ vs centrality for both $\sqrt{s_{NN}}$ = 200 GeV (top) and $\sqrt{s_{NN}}$ = 6.5 TeV (bottom) comparing the Woods-Saxon, Woods-Saxon + Quarks, and $\alpha$ clustering models.}
\label{fig:fig5}
\end{figure}

The effects of different \OO initial-state models on the $v_m\left\lbrace k \right\rbrace$, although not qualitatively significant, should nevertheless be accessible for an analysis of $O(\mathrm{100M})$ events collected in a short \OO run at the LHC.  Additional constraints can be obtained by considering ratios of flow coefficients as functions of centrality, as shown in Figs.~\ref{fig:fig3}-\ref{fig:fig5}.  In this case, both subnucleonic fluctuations and $\alpha$-clustering correlations lead to non-trivial and measurable effects.  Perhaps the strongest effects are visible in the ratio $v_4\left\lbrace 2 \right\rbrace / v_2\left\lbrace 2 \right\rbrace$ (Fig.~\ref{fig:fig4}), where the effects of subnucleonic fluctuations and $\alpha$-clustering tend to act in opposite directions and may even produce detectable non-monotonicity in the corresponding centrality dependences.  This is especially important, given that the beam-energy dependence of the flow ratios contributes to the differences between RHIC and LHC energies in highly non-trivial ways. Quantitatively reproducing both the centrality and $\sqrt{s_{NN}}$ dependences of all flow ratios would therefore place stringent constraints on the importance of subnucleonic fluctuations and $\alpha$-clustering in real-world nuclear collisions, and provides motivation to carry out \OO collisions at both RHIC and LHC energies.

Because $v_4$ appears to be the most promising observable to distinguish $\alpha$-clustering from subnucleonic fluctuations, we also study the quantity $v_4\left\{4\right\}^4$, which is sensitive to the fluctuations of $v_4$ on an event-by-event basis. $v_4\left\{4\right\}^4$ is a particularly interesting observable because hydrodynamic models have so far failed to capture its sign change at the LHC, even for well-understood PbPb collisions \cite{Giacalone:2016mdr,Alba:2017hhe}. In Fig.\ \ref{fig:fig5} we find that at the LHC, there is clear separation in   $v_4\left\{4\right\}^4$ for central collision, which indicates a nice potential for distinguishing between $\alpha$-clustering and subnucleonic fluctuations.  Additionally, these mechanisms produce effects in opposite directions, with $\alpha$-clustering making $v_4\left\{4\right\}^4$ significantly more negative and subnucleonic fluctuations bringing the value of $v_4\left\{4\right\}^4$ close to 0. In contrast, RHIC does not provide a clear signal and it is unlikely that $v_4\left\{4\right\}^4$ could be used to distinguish between our three scenarios. Finally, we have also checked $v_2\left\{4\right\}/v_2\left\{2\right\}$ but found that all three initial conditions produced relatively similar results.

\noindent {\bf Conclusions}. In this work we use ab initio lattice effective field theory calculations of the nuclear structure of $^{16}$O coupled to the state-of-the-art relativistic hydrodynamics description of the Quark Gluon Plasma to determine the possibility of measuring $\alpha$-clustering in relativistic heavy-ion collisions.  We find that LHC energies are better suited to finding $\alpha$-clustering but one must consider ratios of harmonics such as $v_3\{2\}/v_2\{2\}$ and $v_4\{2\}/v_2\{2\}$. Interestingly enough, $\alpha$-clustering suppresses $v_3\{2\}/v_2\{2\}$ and enhances  $v_4\{2\}/v_2\{2\}$ and in all our comparisons subnucleonic fluctuations always has the opposite effect compared to $\alpha$-clustering at LHC energies. Another promising observable is $v_4\{4\}^4$ where very significant differences appear between $\alpha$-clustering and subnucleonic fluctuations between $0-30\%$ centrality at the LHC. In contrast, RHIC has more ambiguous results and appears less likely to be sensitive to $\alpha$-clustering but may be slightly sensitive to substructure.   

While our results for $\mathrm{Kn}$ and $\mathrm{Re}^{-1}$ may be somewhat concerning, this does not immediately rule out the relativistic viscous hydrodynamics picture in small and intermediate systems.  One possible solution may be anisotropic hydrodynamics \cite{Florkowski:2010cf,Martinez:2010sc,Bazow:2013ifa}, re-deriving the hydrodynamic equations of motion in a far-from-equilibrium regime \cite{Denicol:2021wod}, effective transport coefficients \cite{Romatschke:2017vte,Blaizot:2017ucy,Denicol:2018pak,Behtash:2019txb,Denicol:2020eij}, an intermediate stage between initial conditions and hydrodynamics \cite{Kurkela:2018vqr,Kurkela:2018wud} or even considering the $\mathrm{Kn}$ and $\mathrm{Re}^{-1}$ within the Bayesian analysis (and excluding parameter sets with unreasonable results). At the moment we do not look for attractors (originally proposed in \cite{Heller:2015dha}) in our simulations  but leave that for a future work (complications arise in more realistic scenarios with shear and bulk coupled together and a realistic equation of state \cite{Dore:2020jye}). 

\vspace{0.1cm}
\noindent \textit{Acknowledgments}~--~The authors would like to thank  Matthew Sievert    for useful discussions related to this work.  J.N.H. and C.P. acknowledge the support from the US-DOE Nuclear Science Grant No. DE-SC0020633.  D.L. acknowledges support from US-DOE Nuclear Science Grant No. DE-SC0018638, Los Alamos National Laboratory, NUCLEI SciDAC-4 collaboration, and supercomputing resources from the Oak Ridge Leadership Computing Facility and INCITE Award ``Ab-initio Nuclear Structure and Nuclear Reactions'' as well as the J\"{u}lich Supercomputing Centre.

\bibliography{MM}

\end{document}